\documentclass[12pt,a4paper]{article}
\usepackage{graphics}
\usepackage{amssymb}
\begin{document}
\begin{center} {\bf Remarks on the conserved densities of the Camassa-Holm equation}\\
\vskip 0.5 cm
{\small Amitava Choudhuri$^1$, B. Talukdar$^{1a}$ and S. Ghosh$^2$}\\
{\small $^1$\it Department of Physics, Visva-Bharati University, Santiniketan 731235, India}
{\small $^2$\it Patha Bhavana, Visva-Bharati University, Santiniketan 731235, India}\\
\vskip 0.2 cm
{\small \it e-mail : binoy123@sancharnet.in}
\end{center}
\vskip 0.5 cm
\noindent{\bf Abstract}\\
It is pointed out that the higher-order symmetries of the Camassa-Holm (CH) equation are nonlocal and nonlocality poses problems to obtain higher-order conserved densities for this integrable equation (J. Phys. A: Math. Gen. 2005, {\bf 38} 869-880). This difficulty is circumvented by defining a nolinear hierarchy for the CH equation and an explicit expression is constructed for the nth-order conserved density.
\vskip 1.0 cm
The Korteweg-de Vries (KdV) and nonlinear Schr\"{o}dinger (NLS) equations are quasi-linear in the sense that the dispersive behaviour of the solution of each equation is governed by a linear term giving the order of the equation. The dispersion produced is compensated by nonlinear effects resulting in the formation of exponentially localized solitons. In early 1990's, Rosenau and Hyman {[}1{]} introduced a family of nonlinear evolution equations given by
$$K(p,q):\,\,\, u_t+(u^q)_x+(u^p)_{3x}=0\,\,,\,\,\,1<q=p\leq3\,\,.\eqno(1)$$
For the restriction imposed on $p$, the dispersive term of $(1)$ is nonlinear. We shall call evolution equations with nonlinear dispersive terms as fully nonlinear evolution (FNE) equations. The solitary wave solutions of $(1)$  were found to have compact support. That is, they vanish identically outside the finite range. These solutions were given the name compactons. For special values of $q$ and $p$ one can have peaked and cusp-like solutions often called peakon and cuspon {[}2{]}. Besides being nonintegrable, another awkward analytical constraint for the equations in $(1)$ is that they do not follow from an action principle. Almost simultaneously with the work of Rosenau and Hyman, Camassa and Holm {[}3{]} introduced another FNE equation given by
$$u_t-u_{xxt}+3uu_x-2u_xu_{2x}-uu_{3x}=0\,\,.\eqno(2)$$
The Camassa-Holm (CH) equation describes the unidirectional propagation of shallow water waves over a flat bottom. It can also represent the geodesic flow on the Bott-Virasoro group {[}4{]}. An important virtue of $(2)$ is that it follows from an action principle. Further, it admits the Lax pair {[}3{]} and bi-Hamiltonian representation {[}3{]}. The equation is integrable and supports peakon solution.
\par In the recent past a number of works was devoted to construct explicit formulas for conserved densities which are higher in the Camassa-Holm hierarchy {[}5,6,7{]}. One common observation of these studies was that there are inherent difficulties to achieve the goal. The problem arises through the construction procedure to find an appropriate recursive scheme by the use of Hamiltonian duality method as introduced by Fuchssteiner and Fokas {[}8{]}. In this short communication we shall construct a nonlinear hierarchy for the CH equation and make use of a Lagrangian approach to obtain the higher invariants of the CH flow. We note that each member of the derived hierarchy is bi-Hamiltonian. We obtain all results without loss of any mathematical rigor. Despite that, the virtue of our treatment is its simplicity and directness.
\par The CH equation in $(2)$ can be written in the form
$$m_t=\Lambda m_x\eqno(3)$$
with
$$m=u-u_{2x}\,\,\,\,{\rm and}\,\,\,\,\Lambda=\left(-u_x\partial_x^{-1}-\partial_xu\partial_x^{-1}\right)\,\,.\eqno(4)$$
Following Calogero and Degasperis {[}9{]} we identify the pseudo-differential operator $\Lambda$ as a recursive operator that generates a hierarchy of equations according to 
$$m_t=\Lambda^n m_x\,\,,n=0,\,1,\,2,\,....\,\,.\eqno(5)$$
The first member of the hierarchy $(n=0)$ is a linear differential equation
$$u_t-u_{xxt}-u_x+u_{3x}=0\eqno(6)$$
while the second member $(n=1)$ is the CH equation itself. For $(n=2)$ and $(n=3)$ we have
$$u_t-u_{xxt}-6u^2u_x+4uu_xu_{2x}+u_x^3+u^2u_{3x}=0\eqno(7)$$ 
and
$$u_t-u_{xxt}+10u^3u_x-6u^2u_xu_{2x}-3uu_x^3-u^3u_{3x}=0\,\,.\eqno(8)$$
Looking at $(2)$, $(7)$ and $(8)$ we see that as we increase $n$ in $(5)$ we get FNE equation with dispersive terms which are more and more nonlinear. However, the order of the dispersive term remains unchanged. Interestingly, equations similar to those in $(2)$ and $(7)$ can be obtained from $(1)$ for $(p=2)$ and $(p=3)$. The same recursive scheme when applied to the KdV equation according to 
$$u_t=\left(\partial_x^2-4u-2u_x{\partial_x}^{-1}\right)^nu_x\eqno(9)$$
generates equations in the KdV hierarchy. Here the dispersive terms involve higher and higher derivatives as we go along the hierarchy. Thus the CH hierarchy defined in $(5)$ represents a point of departure from the traditional way of introducing hierarchy of quasilinear evolution equations. In spite of this, all equations of the CH hierarchy follow from an action principle. In the following we deal with these and  derive a method to construct expressions for higher invariants in the CH flow.
\par A single evolution equation $u_t=P[u]$, $  u\,\,\epsilon\,\,\mathbb{R}  $ is never the Euler-Lagrange equation of a variational problem {[}10{]}. One common trick to put a single evolution equation into a variational form is to replace $u$ by a potential function
$$u=-w_x,\,\, w=w(x,t)\,\,.\eqno(10)$$
The function $w$ is often called the Casimir potential. We have found that the action functional
$$S=\int\int{\cal L}_n\left(w_t,\,w_x,\,w_{2x},\,w_{3x}\right)dxdt\eqno(11)$$
via the Hamilton's variational principle reproduces the hierarchy of equation defined in $(5)$ with the Lagrangian density written as 
$${\cal L}_n={1\over2}\left(w_x-w_{3x}\right)w_t-{1\over2}\left(w_x\right)^n\left(w_x^2+w_{2x}^2\right)\,\,.\eqno(12)$$
The result obtained from $(12)$ for $n=1$ represents the Lagrangian density of the CH equation {[}11{]}. The Lagrangian density for the first member of the hierarchy in $(5)$ is given by
$${\cal L}_0={1\over2}\left(w_x-w_{3x}\right)w_t-{1\over2}\left(w_x^2+w_{2x}^2\right)\,\,.\eqno(13)$$
By adding a gauge term ${1\over2}\frac{d}{dx}(w_xw_{2x})$, $(13)$ can be written in the form
$${\cal L}_0={1\over2}\left(w_x-w_{3x}\right)w_t-{1\over2}\left(w_x^2-w_xw_{3x}\right)\,\,.\eqno(14)$$
From $(14)$ we write the Hamiltonian density
$${\cal H}_0={1\over2}w_x\left(w_x-w_{3x}\right)\,\,.\eqno(15)$$
In terms of $u$, ${\cal H}_0$ reads
$${\cal H}_0={1\over2}um\,\,.\eqno(16)$$
For $n\geq1$ we construct the Hamiltonian density from $(12)$ such that
$${\cal H}_n={1\over2}\left(-u\right)^n\left(u^2+u_{x}^2\right)\,\,,\eqno(17)$$
while ${\cal H}_0$ in $(16)$ represents the first conserved density of the CH flow. The others are obtained from $(-1)^n{\cal H}_n$. Thus as opposed to the claims in refs. $6$ and $7$ we have obtained a very simple formula for the representation of all constants of the motion and conservation laws for Camassa-Holm equation.
\par It is of interest to note that the hierarchy of nonlinear equations in $(5)$ follows from a Zakharov-Faddeev type Hamiltonian equation {[}12{]}
$$m_t=J_1\frac{\delta {\cal H}_n}{\delta m}\eqno(18)$$
with $J_1=\partial_x(1-\partial_x^2)$ and ${\cal H}_n$ as given in $(16)$ and $(17)$. More significantly, the Camassa-Holm and all associated nonlinear equations as given in $(5)$ are bi-Hamiltonian {[}13{]} in the sense that each of them can be written in the form
$$m_t=J_1\frac{\delta {\cal H}_{n+1}}{\delta m}=J_2\frac{\delta {\cal H}_{n}}{\delta m}\eqno(19)$$
with the second Hamiltonian operator {[}14{]}
$$J_2=\Lambda J_1=-\left(u_x+\partial_xu\right)\left(1-\partial_x^2\right)\,\,.
\eqno(20)$$
\par For integrable models represented by quasilinear evolutions, the Magri recursion scheme {[}13{]} leads to higher-order symmetries and conservation laws which are all local. But the same scheme when applied for the CH equation poses a serious problem {[}5,6,7{]} since in this case the higher-order symmetries are nonlocal and nonlocality does not permit one to construct expressions for higher-conserved densities. We have circumvented this difficulty by defining a nonlinear hierarchy for the CH equation and constructed a simple expression for the nth-order Hamiltonian density. Admittedly one would like to ask: Is there any relation between nonlocality and nonlinearity ? Our answer to this question is fairly straightforward. The dynamics of N identical pairwise interacting quantum particles is governed by a Schr\"odinger equation with a nonlocal potential. In the mean field description this equation reduces to the nolinear Schr\"odinger equation {[}15{]}. Thus the relation between nonlocality and nonlinearity is very transparent for the NLS equation. For other nonlinear equations such association does not appear to be so obvious and one needs further investigation.
\vskip 0.8 cm

This work is supported by the University Grants Commission, Government of India, through grant No. F.10-10/2003(SR).

\vskip 1.2 cm
\noindent{\bf References}\\
\\
{[}1{]} Rosenau P and Hyman J M 1993 Compactons: Solitons with finite wavelength {\it Phys. Rev. Lett.} {\bf 70} 564-67\\
{[}2{]} Rosenau P 1997  On nonanalytic solitory waves formed by a nonlinear dispersion {\it Phys. Lett. A} {\bf 230} 305-18\\
{[}3{]} Camassa R and Holm D 1993 An integrable shallow water equation with peaked solitons {\it Phys. Rev. Lett.} {\bf 70} 1661-4\\
{[}4{]} Misiolek G 1998 A shallow water equation as a geodesic flow on the Bott-Virasoro group {\it J. Geom. Phys.} {\bf 24} 203-8\\
{[}5{]} Fisher M and Schiff J 1999 The Camassa Holm equation: conserved quantities and the initial value problem {\it Phys. Lett. A} {\bf 259} 371-6\\
{[}6{]} Loubet E 2005 About the explicit characterization of Hamiltonians of the Camassa-Holm Hierarchy {\it J. Nonlin. Math. Phys.} {\bf 12} 135-43\\
{[}7{]} Lenells J 2005 Conservation laws of the Camassa-Holm equation {\it J. Phys. A: Math. Gen.} {\bf 38} 869-80\\
{[}8{]} Fuchssteiner B and Fokas A 1981 Symplectic structure, their B\"{a}cklund transformation and hereditary symmetries {\it Physica D} {\bf 4} 47-66\\
{[}9{]} Calogero F and  Degasperis A 1982 {\it Spectral Transform and Solitons} (New York: North-Holland Publising Company)\\
{[}10{]} Olver P J 1993 {\it Application of Lie Groups to Differential Equation} (New York: Springer-Verlag)\\
{[}11{]}Cooper F and Shepard H 1993 Solitons in the Camassa-Holm Shallow Water Equation {\it arXiv patt-sol/9311006}\\
{[}12{]} Zakharov V E and  Faddeev L D 1971 The Korteweg-de Vries equation is a fully integrable Hamiltonian system {\it Func. Anal. Appl.} {\bf 5} 280-87\\
{[}13{]} Magri F  1978 A simple model of the integrable Hamiltoian equation {\it J. Math. Phys.} {\bf 19} 1156-78\\
{[}14{]} Ref.{[}9{]} page, 453\\
{[}15{]} Whitham G B 1974 {\it Linear and Nonlinear Waves} (New York)\\
\end{document}